\documentclass[usenatbib,usegraphicx]{mn2e}
\usepackage{aas_macros}
\usepackage{natbib}
\usepackage{amsmath,amssymb,amsfonts,bm}
\usepackage{subfigure}
\usepackage{longtable}
\usepackage{graphicx}
\usepackage{graphics}
\usepackage{keyval}
\usepackage{trig}
\usepackage{float}
\usepackage{epsfig}
\usepackage[dvipsnames]{color}
\usepackage{calc}
\usepackage{url}
\usepackage{hyperref}
\def\beq{\begin{equation}}
\def\eeq{\end{equation}}
\def\bey{\begin{eqnarray}}
\def\eey{\end{eqnarray}}
\def\Myr{\, {\rm Myr} }
\def\Gyr{\, {\rm Gyr} }

\def\pc{\, {\rm pc} }

\def\kpc{\, {\rm kpc} }
\def\mpc{\, {\rm Mpc} }
\def\msun{\rm M_\odot}

\def\kms{\, {\rm km \, s}^{-1} }

\def\a0{$a_0$}

\def\ge{{\bf g}_{\rm ext}}

\def\mvir{M_{\rm vir}}
\def\rvir{r_{\rm vir}}

\def\mssq{\rm m~s^{-2}}
\begin{document}
\title{Galactic rotation curves, the baryon-to-dark-halo-mass relation and space$-$time scale invariance}
\author[]{Xufen Wu$^{1,*}$, Pavel Kroupa$^{2,*}$  \\
$^{1}$ Deparment of Astronomy, University of Science and Technology of China,
Jinzai Road 96, 230026, Hefei, China\\
$^{2}$ Helmholtz-Institut f\"{u}r Strahlen-und Kernphysik,
Universit\"{a}t Bonn, Nussallee 14-16, D-53115 Bonn, Germany\\
$^{*}$ Email: xwu@astro.uni-bonn.de; pavel@astro.uni-bonn.de\\
 } \maketitle

\begin{abstract}
  Low-acceleration space$-$time scale invariant dynamics (SID,
  \citealt{Milgrom2009c}) predicts two fundamental correlations known
  from observational galactic dynamics: the baryonic Tully-Fisher
  relation (BTFR) and a correlation between the observed mass
  discrepancy and acceleration (MDA) in the low acceleration regime for 
  disc galaxies. SID corresponds
  to the deep MOdified Newtonian Dynammics (MOND) limit. 
  The MDA data emerging
  in cold/warm dark matter (C/WDM) cosmological simulations disagree
  significantly with the tight MDA correlation of the observed
  galaxies. Therefore, the most modern simulated disc galaxies, which
  are delicately selected to have a quiet merging history in a standard
  dark-matter-cosmological model, still do not represent the correct
  rotation curves.  Also, the observed tight correlation contradicts
  the postulated stochastic formation of galaxies in low-mass DM
  haloes. Moreover, we find that SID predicts a baryonic to apparent
  virial halo (dark matter) mass relation which agrees well with the
  correlation deduced observationally assuming Newtonian dynamics to
  be valid, while the baryonic to halo mass relation predicted from
  CDM models does not. The distribution of the observed ratios of
  dark-matter halo masses to baryonic masses may be empirical evidence
  for the external field effect, which is predicted in SID as a
  consequence of the forces acting between two galaxies depending on
  the position and mass of a third galaxy. Applying the external field
  effect, we predict the masses of galaxies in the proximity of the
  dwarf galaxies in the Miller et al. sample. Classical
  non-relativistic gravitational dynamics is thus best described as
  being Milgromian, rather than Newtonian.
\end{abstract}

\section {Introduction}
The currently widely accepted understanding of gravity is based
entirely on the empirical law derived by Newton. Newton based his
derivation on a number of observations, the central ones being the
laws of planetary motion proposed by Kepler. \cite{Einstein1916}
revolutionized our concept of gravitation as being not a force but an
effect due to space$-$time curvature. Einstein's field equation is in
turn based on Newton's law derived on the Solar System scale: at the
time of Einstein's proposal in 1916, galaxies had not been discovered
to be what we know them to be nowadays. Thus, when applied to galaxies
and cosmological scales, the Einsteinian/Newtonian law of gravity
constitutes an extrapolation by many orders of magnitude in spatial
and acceleration scale beyond the gravitational systems known in
1916. The observation that the rotation curves of galaxies deviate
from the expected Keplerian decline by being essentially flat at large
radii, $r$ \citep{Rubin_Ford1970,Bosma1981,Rubin+1982}, i.e. that the
circular velocities of galaxies obey $v_{circ}\approx
\mathrm{constant}$, is therefore a very major discovery. This behavior
violates Newton's empirical law of gravity that predicts a Keplerian
fall off of the circular velocity, $v_{circ}\propto r^{-1/2}$, in the
outer regimes of galaxies.  It is not entirely surprising that
Einsteinian/Newtonian gravity breaks down at those scales. However,
today the popular interpretation of this discrepancy is to assume
  Newtonian/Einsteinian gravitation to be valid and to postulate the
existence of cold (C) or warm (W) dark matter (DM) particles which
make up the mass-discrepancy when rotation curves are interpreted in
terms of Einsteinian/Newtonian gravitation. However there exists no
experimental evidence for the existence of additional (e.g., dark
matter) particles beyond those predicted or contained within the
standard model of particle physics despite a highly significant effort
for finding them \citep[see e.g. the recent press release from the
direct search for dark matter particles with the Large Underground
Xenon dark matter detector,][]{LUX2013, Kroupa2014}.  Therefore it is
important to test alternative gravities, amongst which Milgromian
dynamics\footnote{Milgromian dynamics, often referred to as Modified
  Newtonian Dynamics (MOND), is briefly introduced in
  Appendix~\ref{MOND}, but here focus is on the deep-MOND or
  weak-acceleration limit, which is the regime of scale-invariant
  dynamics (SID, Sec.~\ref{STSI}).} \citep{Milgrom1983a} is the most
promising one.

On the scale of galaxies, a mass-discrepancy$-$acceleration (MDA) correlation has
been predicted by \citet{Milgrom1983a}: there is an exact correlation
between the mass discrepancy (i.e., the amount of unseen additional
mass needed when interpreting the observed motions within the
Newtonian dynamics framework) and the acceleration deduced from the
orbits at all radii observed in galaxies.  The mass discrepancy,
$\frac{M_{dyn}(<r)}{M_{\rm b}(<r)} \propto
\left[\frac{v(r)}{v_{\rm b}(r)}\right]^2$, since it is well known that

\bey \label{vcirc2} 
v(r)^2 &\equiv & g(r) r \simeq \frac{GM_{dyn}(<r)}{r}, \\
\label{vcircb2} 
v_{\rm b}(r)^2 &\equiv & g_{\rm N}(r) r \simeq \frac{GM_{\rm b}(<r)}{r},
\eey 

\noindent where $M_{dyn}(<r)$ is the dynamical mass (i.e., the total
Newtonian mass) within $r$, $v(r)$ is the total observed or actual
rotation speed at $r$, $v_{\rm b}(r)$ is the rotation speed at $r$
contributed only from the baryons assuming Newtonian dynamics (note
that ``$\simeq$'' in Eqs. \ref{vcirc2}-\ref{vcircb2} becomes an
equality when the system is spherically symmetric).

This MDA correlation has been quantified empirically by
\citet{Sanders1990} and more recently by \citet{McGaugh2004} for a
sample of 74 disc galaxies. This correlation is extended to the Solar
System scale by \citet[][their
fig. 4]{Famaey_McGaugh2012}. \citet{Trippe2013} fitted the MDA relation using McGaugh (2004) data in a massive graviton model, which is equivalent to MOND with a simple interpolating function.
\citet{Scarpa2006} tested this
correlation for a sample of over 1000 pressure-supported systems from
globular clusters to rich clusters of galaxies. More precisely,
\citet{Scarpa2006} studied the correlation between $g$ and $g_{\rm N}$
instead of the MDA correlation. However, these two correlations,
though formally different, are fully
equivalent. \citet{Tiret_Combes2009} examined the MDA correlation for
a sample of 43 galaxies including early- and late-type galaxies.  

The prediction by \cite{Milgrom1983a} of the MDA correlation and its
subsequent empirical confirmation offer detailed tests of theories of
galaxy formation and dynamics.  One of the major questions addressed
with this contribution is whether galaxies simulated in the W/CDM
cosmological frameworks also reproduce the observed MDA
correlation. This is a timely question to ask, because recently there
have been clams that disc galaxies can form in the DM models and that
these galaxies also resemble real galaxies.

Noteworthy is that the deep Milgromian limit (i.e., dynamics in the
weak field regime where $g\ll a_0$; here Milgrom's constant
$a_0\approx 1.2\times10^{-10}$m/s$^2$ can be derived from space$-$time
scale invariance and appears to be a constant of nature) can be
described extremely well with low-acceleration space$-$time
scale-invariant dynamics (hereafter SID, Sec.~\ref{STSI}), as
originally pointed out by \citet{Milgrom2009c}.  With this
contribution we revisit pure SID. It is shown, using simple arguments,
how phantom (ie. unreal) dark matter haloes, the BTFR and the MDA
correlation emerge naturally within SID for an observer who interprets
observations in terms of Newtonian dynamics. Therewith this further
affirms the Milgrom's and Bekenstein's extension of effective gravity
beyond the 1916 version as being realistic, in contrast to the
introduction of C/WDM particles which are speculated to exist outside
the standard model of physics.

The Strong Equivalence Principle (SEP) is violated in SID because of
external fields \citep{Milgrom1983a,BM1984}. This violation is a
direct outcome of the truncation of the phantom dark matter haloes as
can be demonstrated nicely in the pure SID regime. {\it This means
  that in SID, the force acting between two galaxies depends on the
  position and mass of a third galaxy}.  The violation of SEP is one
of the most fundamental differences between SID/MOND and W/CDM. The
violation of the SEP means that SID/MOND cannot be obtained by taking
the non-relativistic limit of Einstein's GR, which is known to be
built on the basis of SEP. The internal dynamics of a system embedded
in an external field depends on both the internal and the external
gravitational fields \citep{Milgrom1983a,BM1984}. Therefore, it is
important to study the effects of the external fields for real
galaxies embedded in clusters of galaxies, and of satellite galaxies,
to improve our constraints on the validity of SID and thus of
Milgromian dynamics.

SID is discussed in Sec.~\ref{STSI}.  The MDA correlation is shown to
directly and immediately emerge from SID.  The connection between SID
and MOND is discussed in Appendix~\ref{MOND}.  We then visit in
Sec.~\ref{cosmology} the standard interpretation of the MDA
correlation in terms of C/WDM and consult the most advanced
astrophysical models of disc galaxy formation, which claim to explain
disc galaxies readily.  The baryonic-to-dark-matter-halo mass relation
is studied in Sec.~\ref{masses} for the SID prediction and for the
C/WDM simulations, and a comparison to the observations is
provided. The external field effect predicts a truncation of the
phantom dark matter haloes in SID. This is studied in Sec.~\ref{efe},
where possible observational evidence for this important truncation is
documented. We conclude with Sec.~\ref{conc}.

\section{space$-$time scale invariant dynamics and the MDA correlation}\label{STSI}

In this section we revisit low-acceleration space$-$time scale-invariant
dynamics (SID) raised by \cite{Milgrom2009c} concerning the deep-MOND
limit. The reader is reminded that most of the universe is in the SID
regime \citep{Kroupa2014}. It is also shown that in the low acceleration regime the MDA correlation follows from SID
and the BTFR also immediately follows from SID. We thereby stress again that
is a remarkable fact that such a simple symmetry, discovered by
\cite{Milgrom2009c}, leads to such profound and correct reproduction
of the most important laws of observed galactic dynamics.  It also
follows from SID using simple arguments that apparent
(i.e. non-particle) ``dark matter haloes'' arise simply and directly,
as is shown in sec.~2.1 in \cite{Wu_Kroupa2013} and this is applied to
the external field effect in Sec.~\ref{pdmmass}--\ref{virial} to
demonstrate that the effective gravitating masses of galaxies depend
on the position and mass of neighboring galaxies.

Consider a space$-$time scale invariance of the equations of motion
under the consideration of the transformation in Minkowsky space
(\citealt{Milgrom2009c}; see also
\citealt{Milgrom2014a,Milgrom2014b}),

\beq\label{SIDeq} 
(t,{\bf r}) \rightarrow (\lambda t,\lambda {\bf r}), 
\eeq 

\noindent where $t$ and ${\bf r}=(x,y,z)$ are time and Cartesian
coordinates, respectively, and $\lambda$ is a positive number. The
Newtonian gravitational acceleration for a spherically symmetric
system,

\beq 
g_{\rm N} = \frac{GM_{\rm b}}{r^2},
\eeq 

\noindent then transforms as $g_{\rm N} \rightarrow \lambda^{-2}g_{\rm N}$,
whereas the kinematical acceleration, $g\equiv {\rm d} \dot{x} / {\rm
  d} t$, scales as $g \rightarrow \lambda^{-1}g$. Here $M_{\rm b}(<r)$ is
the enclosed baryonic mass within $r$. As a result, the Newtonian
gravitational acceleration and the kinematical acceleration scale
differently under Eq.~\ref{SIDeq}. Linking purely gravitational
interactions to symmetries such as defined in Eq. \ref{SIDeq} suggests
deeper physics and constitutes a motivation for viewing MOND as much
more than a mere phenomenological description of galactic dynamics
\citep{Milgrom2009a}.  In order to assure that both, the gravitational
and the kinematical accelerations scale symmetrically under
Eq.~\ref{SIDeq}, that is, in order to maintain the invariant symmetry,
the gravitational acceleration, $g$, has to scale proportionally to
$g_{\rm N}^{1/2}$. In order to obtain the correct dimension, a constant with
the unit of acceleration, needs to be introduced. This constant is
referred to as $a_0$, such that

\beq\label{SIDg}g=(a_0g_{\rm N})^{1/2},
\eeq

\noindent i.e. $g^2=a_0g_{\rm N}$. Thus $g=(GM_{\rm b} a_0)^{1/2}/r$, and the
circular velocity, which follows from the centrifugal acceleration
$g=v^2/r$, is

\beq\label{vc} v = (GM_{\rm b}a_0)^{1/4}={\rm constant},
\eeq

\noindent which is exactly the BTFR
\citep{Milgrom1983a,McGaugh+2000,Milgrom2009c,Famaey_McGaugh2012,
  Milgrom2014a}. We refer to gravitational dynamics which thus
conforms to low-acceleration scale-invariance (Eq. \ref{SIDeq}) as
low-acceleration scale-invariant dynamics (SID). SID beautifully
reproduces the deep MOND equations of motion. It is rather remarkable
that such a simple principle as SID and discovered by Milgrom leads to
one of the most important scaling relations which real galaxies are
observed to obey. Note that Eq.~\ref{vc} implies that each baryonic
galaxy is surrounded by a logarithmic non-particle (and thus phantom)
dark matter halo potential, which is however not a real halo as it is
only evident if Newtonian dynamics is applied to the galaxy. If SID is
true, then a Newtonian observer would thus deduce that each baryonic
galaxy is surrounded by a phantom DM halo the mass of which is
proportional to radial distance (Eq.~\ref{mpdmh} below).

Under SID and due to Eq. \ref{vcirc2}$-$\ref{vcircb2}, for a spherical
system, the mass discrepancy,

\beq\label{vvb_g} 
\left(\frac{v}{v_{\rm b}}\right)^2=\frac{g}{r} \cdot
\frac{r}{g_{\rm N}}=\frac{g}{g_{\rm N}}, \eeq 

\noindent becomes $\frac{(GM_{\rm b}
  a_0)^{1/2}}{GM_{\rm b}/r}=\left(\frac{a_0}{g_{\rm N}}\right)^{1/2}$. Thus SID
immediately implies a simple relation between the mass discrepancy and
the baryonic Newtonian acceleration under the invariance
transformation,

\beq\label{mdg}
\left(\frac{v}{v_{\rm b}}\right)^2=\left(\frac{a_0}{g_{\rm N}}\right)^{1/2}.
\eeq

\begin{figure}{}
\begin{center}
  \resizebox{9.cm}{!}{\includegraphics{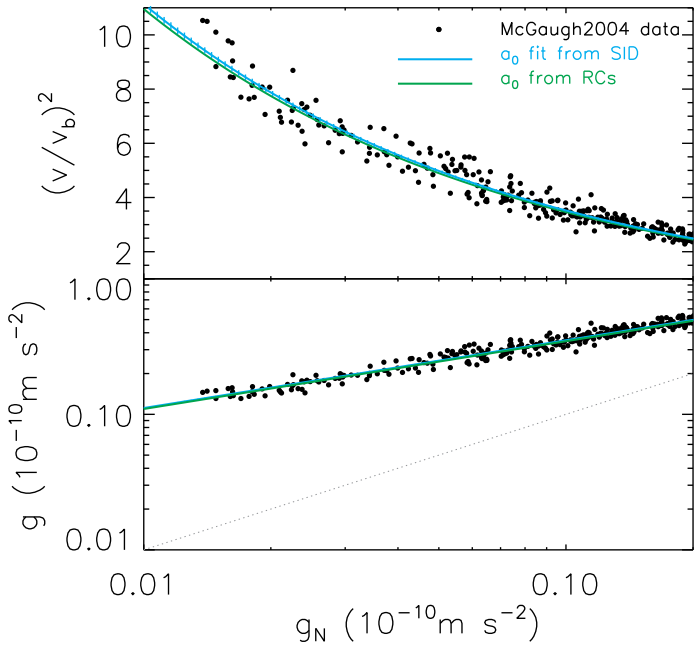}}
  \makeatletter\def\@captype{figure}\makeatother \caption{Upper panel:
    The mass discrepancy $(v/v_{\rm b})^2$ versus baryonic acceleration
    $g_{N}$ in the weak field ($g_{\rm N}< 0.2 a_0 \ll a_0$) regime. The
    black points are the observed
    mass-discrepancy--baryonic-acceleration data by
    \citet{McGaugh2004}, and the cyan curve is the best fit of the MDA
    correlation for these data (see Eq. \ref{mdg}), where $a_0=1.24
    \times 10^{-10} m~s^{-2}$. The green curve is the MDA correlation
    with $a_0$ determined from observations of galactic rotation
    curves \citep[see Sec. \ref{STSI},][]{Begeman+1991}. Lower panel:
    the kinematical acceleration versus Newtonian gravitational
    acceleration in the weak field regime. The observed data show a
    tight correlation between $g$ and $g_{\rm N}$. The faint dotted
      line shows the $g=g_{\rm N}$ relation, which is not followed by the
      observational data.
     }\label{weakfield}
\end{center}
\end{figure}

This function (Eq.~\ref{mdg}) is plotted in the upper panel of
Fig~\ref{weakfield}, where it is compared with the observational data
in the weak-field regime ($g_{\rm N}<0.2\times 10^{-10}\,$m/s$^2$).
The lower panel of this figure also shows the observationally deduced
acceleration, $g$, in dependence of $g_{\rm N}$. A value for $a_0$ can
be obtained by fitting Eq. \ref{mdg} to the data points in
Fig. \ref{weakfield} (cyan curve). A Levenberg-Marquardt fit to the
observed MDA data is shown in the upper panel of
Fig. \ref{weakfield}. {\it It follows that pure SID constitutes an
  excellent description of the observational data for $a_0=1.24\pm
  0.03 \times 10^{-10}\mssq = 3.90 \pm 0.01 \pc \Myr^{-2}$} (within
$1\sigma$ confidence level) for data points within the above mentioned
weak field regime.

SID is broken near gravitating masses whenever $g$ approaches $a_0$
from below such that gravitational dynamics becomes Newtonian.  A
connection between the Newtonian regime and the SID regime is required
to study the kinematic acceleration in the transitionary regime where
$g\simeq a_0$. To study the kinematics of galaxies in the full regime
of acceleration, a transition function $\mu$ is introduced by
\citet{Milgrom1983a}, yielding the full MOND description of galactic
dynamics (Appendix~\ref{MOND}).  The kinematic acceleration, ${\bf
  g}$, is transformed from the Newtonian acceleration, ${\bf g}_N$,
through

\beq\label{interpol}
{\bf g}_N=\mu(|{\bf g}|/a_0){\bf g},
\eeq 

\noindent with the help of the $\mu(|{\bf g}|/a_0)$
function. \footnote{Note that this is, from the procedural point of
  view, equivalent to \citet{Planck1901} introducing the constant
  $\hbar$ as an auxiliary constant (``Hilfsgr{\"o}{\ss}e'' in German)
  to describe the transition between the low-energy Rayleigh-Jeans
  black body spectrum and the downturn towards high-energies observed
  for black body radiators. The quantisation of energy had not been
  realised to be the underlying physics until more than 25 years
  later. } Several forms of the $\mu(|{\bf g}|/a_0)$ function have
been proposed by \citet{Milgrom1983a,Milgrom1999} and
\citet{Bekenstein2004,Famaey_Binney2005,zhao2008}. We use two of the
most popular interpolating functions: the `simple'$~\mu$-function and
the `standard'$~\mu$-function,

\bey\label{mu} 
\mu(x)&=&\frac{x}{1+x},\quad {\rm ~~`simple' ~\mu},\nonumber\\
\mu(x)&=&\frac{x}{\sqrt{1+x^2}},\quad {\rm ~~`standard'~ \mu},\\
x&\equiv &|{\bf g}|/a_0. \nonumber 
\eey

We note here that the complete Milgromian description of classical
gravity has been shown to have a Lagrangian formulation \citep{BM1984}
such that this theory is energy and angular momentum conserving.  For
a spherically symmetric, cylindrically symmetric or axi-symmetrically
system, the solution of the Lagrangian formulation takes the
simplified form, Eq. \ref{interpol} \citep{BM1984}. The standard form
of the $\mu$-function can be associated with quantum mechanical
processes in the vacuum (\citealt{Milgrom1999}, see also Appendix~A in
\citealt{Kroupa+2010}).

Thus a relation between the mass discrepancy and the acceleration over
the full classical regime is determined by the interpolating
$\mu$-function: 

\beq\label{vvb_mu} \left(\frac{v}{v_{\rm b}}\right)^2=\frac{g\cdot r}{g_{\rm N}
  \cdot r} \simeq \frac{1}{\mu(x)}.
\eeq

This indicates that the formation of galaxies cannot be stochastic,
since otherwise the observational MDA data would have a much wider
spread. The MDA correlation conflicts with the requirement of
$\Lambda$CDM cosmological simulations that galaxy formation in
low-mass DM haloes must be stochastic \citep{Boylan-Kolchin+2011}.  In
such a speculative stochastic galaxy formation model, the C/WDM halo
mass ceases to be correlated with its baryonic/luminous galaxy. A
further discussion of the MDA correlation for $\Lambda$CDM-simulated
galaxies can be found in Sec.~\ref{cosmology}. More quantitative
studies of the different forms of the $\mu$-function will also be
presented in Sec. \ref{cosmology}.

\section{Standard cosmological models and the MDA correlation}\label{cosmology}

Despite the observations indicating rather convincingly that
gravitation in the classical regime is Milgromian (Sec.  \ref{STSI}),
it is more popular to describe galactic dynamics by postulating
Newtonian gravity to be valid as a major extrapolation from the Solar
System scale to the scale of galaxies plus the existence of
dynamically significant C/WDM particles which are neither
  described by nor contained within the otherwise highly successful
  standard model of particle physics
\citep{Blumenthal+1984,Davis+1985}. The resulting standard
cosmological model, the $\Lambda$C/WDM model, has been subject to
significant testing \citep{Kroupa+2010,Kroupa2012, Kroupa2014}. One of
the major problems for the $\Lambda$CDM model is that the merging
history of each major dark matter halo makes the formation of disc
galaxies highly problematical and until now not convincingly
successful.

In the past two decades, the formation of disc galaxies has been
extensively investigated by means of standard $\Lambda$CDM
cosmological simulations \citep{Katz_Gunn1991,Navarro_Benz1991,
  Navarro_Steinmetz1997, Weil+1998, Abadi+2003,Piontek_Steinmetz2011,
  Hummels_Bryan2012, Agertz2011, Guedes2011,Aumer+2013,
  Marinacci+2013}. The rotation curve has attracted particular
interest since it is one of the most important tools to examine the
reality of the simulated disc galaxies. It had been shown that the
simulated disc galaxies have unrealistic rotation curves with sharp
peaks at their centres declining at larger radii
\citep{Navarro_Benz1991,Navarro_Steinmetz1997,Weil+1998,
  Abadi+2003,Piontek_Steinmetz2011,Hummels_Bryan2012}, which disagrees
with the observations \citep{Rubin+1985}. This is a result of the
simulated galaxies having too little angular momentum because of their
merging history which is inherent in the standard model by virtue of
dynamical friction between the dark matter haloes, and because the
gaseous baryons lose orbital energy by being compressed and by
dissipating kinetic energy as they fall into the deep potential well
of a dark matter halo.

Recent studies claim that disc galaxies with `realistic' shapes of
rotation curves can form in cosmological simulations with gas by using
a subgrid model which enhances the efficiency of stellar feedback
\citep{Governato+2010, Agertz2011, Guedes2011, Stinson+2013,
  Aumer+2013, Marinacci+2013}. Note that the feedback processes of
model disc galaxies are fitted artificially to ``agree with'' the
observations, and these feedback processes are neither natural results
of the cosmological simulations nor predictions of dark
matter. This also applies to the recent results from the Illustris project: the feedback applied there is unphysical and the BTFR comes out incorrectly \citep{Vogelsberger+2014,Kroupa2014}. Another criticism of the feedback processes can be found in the most recent EAGLE project  publication \citep{Schaye+2014}. However, in spite of the artificiality of feedback processes,
we would like to know: how realistic are these model disc galaxies?

If the simulated disc galaxies represent real galaxies, there should
be a MDA correlation in these galaxies, and such relations have to
agree with the empirical data. We here study the MDA correlation of
simulated disc galaxies
\citep{Agertz2011,Guedes2011,Aumer+2013,Marinacci+2013}, which are
claimed to be more realistic than obtained in previous work and which
are taken to demonstrate that the standard cosmological model can,
after all, account for the observed galaxies.

Apart from the problems of rotation curves, there are other
difficulties with the $\Lambda$CDM cosmology simulations on galactic
scales, such as the cusp vs core problem \citep{Springel+2008} and the
missing satellites problem (or more correctly, the satellite
over-prediction problem) \citep{Klypin1999,Moore+1999} and many other
failures to account for data
\citep{Kroupa+2010,Kroupa2012,Famaey_McGaugh2012, Kroupa2014}. To
overcome these difficulties, an alternative dark matter particle has
been proposed, a thermal relic WDM particle with a keV mass scale
\citep{Colin+2000, Bode+2001, Gao_Theuns2007, Schneider+2012}.  The
time scale for structure formation in the WDM cosmology is longer than
that in the CDM cosmology, and the amount of substructures is
decreased. The pioneering work on the formation of disc galaxies in
WDM cosmological models is carried out by \citet{Herpich+2013},
according to which the disc galaxies have less centrally concentrated
stellar profiles in improved agreement with real galaxies.

We here also study the MDA correlation of these galaxies.
The circular velocities of galaxies contributed from baryons, $v_{\rm b}$,
and from dark matter, $v_{DM}$, are taken from the above simulations. The
mass discrepancy is \beq\label{massdiscrepancy}
\left[\frac{v(r)}{v_{\rm b}(r)}\right]^2
=\frac{v_{\rm b}^2(r)+v_{DM}^2(r)}{v_{\rm b}^2(r)} . \eeq The Newtonian
acceleration (from baryons only) can be calculated from the circular
velocity, \beq\label{vcn} g_{\rm N}(r)=v_{\rm b}^2/r.\eeq

\begin{figure}{}
\begin{center}
  \resizebox{9.cm}{!}{\includegraphics{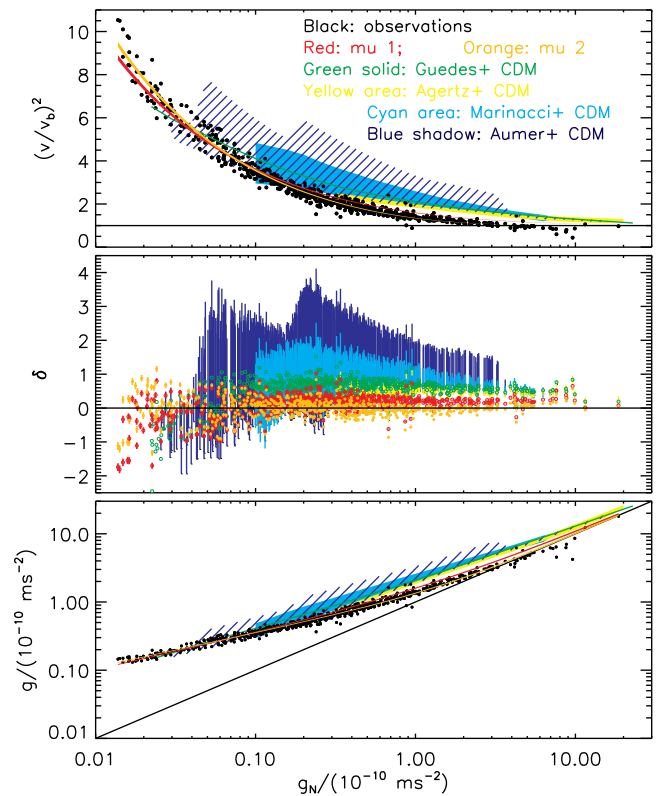}}
  \makeatletter\def\@captype{figure}\makeatother \caption{Upper panel:
    the mass discrepancy $(v/v_{\rm b})^2$ versus baryonic acceleration,
    $g_{\rm N}$, from weak to strong fields. The black points are the
    observed MDA data by \citet{McGaugh2004} for a sample of 74 disc
    galaxies, including both dwarf and Milky Way-scale spiral
    galaxies. The red and orange areas are the correlations predicted
    by Milgromian dynamics with different interpolating functions, red
    for simple $\mu$ and orange for standard $\mu$ (Eq. \ref{mu}).
    The values of $a_0$ and the corresponding error are listed in the
    $6_{th}$ and $7_{th}$ columns of Table \ref{sign}.  The green
    curve shows the mass discrepancy-acceleration relation of
    simulated Milky Way-scale galaxies by \citet{Guedes2011}. The
    coloured shadow areas are the mass discrepancy-baryonic
    gravitational acceleration data for simulated galaxies from cold
    dark matter cosmological simulations \citep[upper
    panel][]{Agertz2011,Marinacci+2013,Aumer+2013}. Middle panel: the
    difference of the mass discrepancy between CDM models and observed
    galaxies, $\delta$, which is defined in Eq. \ref{delta_vvb}.  For
    the simulated and observed galaxies $g_{\rm N}$ is computed from
    Eq. \ref{vcn}.  Note that Milgromian dynamics with the standard
    $\mu$ function (orange curve) is an excellent description of the
    data (black dots in upper panel), the differences of which to the
    Milgromian dynamics relation are plotted as orange dots. The red
    dots are the differences between the observed (black points) and
    the Milgromian dynamics curve with the simple $\mu$ function.  The
    apparent larger scatter in the orange $\delta$ values at small
    $g_{\rm N}$ is expected for a constant (small) dispersion of data around
    the Milgromian dynamics curve. Lower panel: the kinematic-luminous
    acceleration relation of the CDM models and the observed galaxies
    (symbols and colors as in the upper panel). The black line
    corresponds to $g=g_{\rm N}$.  }\label{disp_cdm}
\end{center}
\end{figure}

\subsection{Disc galaxies from CDM cosmological simulations}\label{cdm}
In the dark-matter approach, there are two primary
formation scenarios for disc galaxies. The first one was studied by
\citet{Eggen+1962,Samland_Gerhard2003} and \citet[][model
S1]{Sommer-Larsen+2003}, which is to form a disc galaxy in a growing
dark matter halo by accreting gas (and also dark matter in
\citealt{Samland_Gerhard2003}) in the absence of mergers of dark
haloes. This work was generally not accepted by the community because
it lacked a ``realistic'' merging history. The second scenario is to
form disc galaxies through accreting a large number of satellite
galaxies \citep[so called minor mergers,
e.g.,][]{Bullock_Johnston2005,Moore+2006}. The latter scenario was
more favoured since it is consistent with the hierarchical assembly
of CDM haloes in cosmological simulations \citep{Helmi2008}. However,
it is difficult to produce disc-dominated and bulgeless galaxies
from the simulations of cosmological mergers, while there is a large
fraction ($\approx 70\%$) of edge-on disc galaxies which are
bulgeless or disc-dominated in observations \citep[in a complete and
homogeneous sample of 15127 edge-on disc galaxies in the SDSS data
release 6,][]{Kautsch2009}. HST photometry and Hobby-Eberly
Telescope (HET) spectroscopy of giant Sc-Scd galaxies
\citep{Kormendy+2010} also shows that more than $50\%$ of a sample
of 19 galaxies are bulgeless galaxies, which challenges the picture
of galaxy formation by hierarchical merging. Baryonic feedback
processes have been studied so as to save the second scenario
\citep[e.g.,][]{Okamoto+2005,Piontek_Steinmetz2011,Scannapieco+2012}. However
it has been shown that the angular momentum problem of the simulated
disc-dominated galaxies and bulgeless galaxies cannot be solved by
adding feedback process and by increasing the numerical resolution
of the simulations \citep[e.g.][]{DOnghia+2004,Piontek_Steinmetz2011}.

In a more recent study based on the Millennium-II simulations
\citep{Fakhouri+2010} about $31\%$ of the Galaxy-scale haloes have
experienced a major merger since $z=1$ (corresponding to a look-back
time of about $7-8\Gyr$), and the fraction of major mergers rises to
$69\%$ since $z=3$ (corresponding to a look-back time of about
$11-12\Gyr$). Moreover, in cosmological simulations, over the last
$10\Gyr$ for Galaxy-scale haloes with a mass of $\approx 10^{12}\msun
h^{-1}$, $95\%$ of them have undergone a minor merger by accreting a
subhalo with mass $>5\times 10^{10} \msun h^{-1}$, and $70\%$ of them
have accreted a subhalo with mass $>10^{11} \msun h^{-1}$
\citep{Stewart+2008}. Therefore mergers are very common for
Milky-Way-scale haloes in cosmological simulations.  The simulations of
major mergers (with equal mass galaxy pairs) show that the disc can be
completely disrupted and that the remnants of such mergers become
early-type galaxies \citep[elliptical galaxies or bulge-dominated
galaxies,][]{Toomre1977,Cox_Loeb2008}, and minor mergers (with a mass
ratio $10:1$) also lead to growth of the bulge and thickness of the
disc \citep[e.g.,][]{Walker+1996,Naab_Burkert2003,Younger+2007,
  Kazantzidis+2009}.  Thus, the very large fraction of observed
bulgeless disc galaxies ($70\%$ in edge-on disc galaxies) is
inconsistent with the high incidence ($>70\%$) of significant mergers,
a point also emphasized by \citet{Kormendy+2010}.

More recently, a series of new models using smoothed-particle
hydrodynamics (SPH) simulations \citep{Agertz2011, Guedes2011,
  Aumer+2013,Marinacci+2013} claimed that more realistic disc galaxies
are formed in a quiet merger history scenario. In such a scenario no
mergers with a mass ratio of the substructure and host galaxy larger
than $1:10$ are allowed at low redshift. These authors have thus,
essentially, returned to the previously discussed models by
\citet{Samland_Gerhard2003} which, however, had been criticized as
lacking cosmological realism by being void of mergers.  Thus these
galaxies are selected from the unlikely fraction of Milky-Way-scale
haloes in standard cosmological simulations. It has been claimed that
the rotation curves of these simulated galaxies have more reasonable
shapes without sharp central peaks. To test these simulated galaxies
more quantitatively and in more detail, we now study their MDA
correlation, and then compare the theoretical relation with that
extracted from the observed galaxies.

The relation is shown in the upper panel of Fig. \ref{disp_cdm}, where
the green line, coloured shadows and areas represent the relations of
disc galaxies in the aforementioned simulations, respectively.  A
correlation between the mass discrepancy at radius r,
$\left[\frac{v(r)}{v_{\rm b}(r)}\right]^2$, and the baryonic Newtonian
acceleration at the same radius, Eq. \ref{vcn}, does exist in the
simulated galaxies. However, despite the existence of such a
correlation, the simulated galaxies are not consistent with
observations.  The MDA correlation obtained from the simulated disc
galaxies \citep{Agertz2011,Guedes2011,Aumer+2013,Marinacci+2013} lies
significantly above the empirical relation. Among the above
simulations, there is one Milky Way like disc galaxy modeled by
\citet{Guedes2011}, in which the trend of the mass
discrepancy-acceleration relation is different from that obtained from
the observed data points: with decrease of the acceleration, the
increase of mass discrepancy is slower in the \citet{Guedes2011} model
galaxy. Note that the disc galaxies in
\citet{Agertz2011,Marinacci+2013} and \citet{Guedes2011} are obtained
from re-simulations of haloes with a quiet merger history, i.e., from
haloes without major mergers at low redshift.  The disc galaxies in
\citet{Aumer+2013} are selected from haloes both with and without low-z
major mergers. There is a much wider spread and a larger deviation
from the observational data for these \citet{Aumer+2013} galaxies
compared to other samples, and this could be an effect of the low-z
mergers.

Although there is an overlap of the MDA correlations from simulations
and observations in the weak field regime where $g_{\rm N}<0.1a_0$, the
majority of MDA correlations predicted from the re-simulated disc
galaxies are inconsistent with observations.  More concretely, for a
given enclosed baryonic mass $M_{\rm b}$, the mass discrepancy is always
over-predicted.

Therefore, considering the detailed rotation curves of galaxies
simulated in a CDM universe, the simulated galaxies do not agree with
the observed ones.  The host dark matter haloes have to bear very quiet
merger histories. Since the vast majority of local galaxies ($72\%$
spiral galaxies and $15\%$ S0 galaxies, overall $87\%$) are disc
galaxies \citep{Delgado-Serrano+2010}, this generates another severe
issue that cannot be resolved within the dark matter models.

The middle panel of Fig. \ref{disp_cdm} shows the difference of the
MDA correlation between the simulated
$\left[\frac{v(g_{\rm N})}{v_{\rm b}(g_{\rm N})}\right]^2_{\rm CDM} $ and the observed
galaxies $\left[\frac{v(g_{\rm N}}{v_{\rm b}(g_{\rm N})}\right]^2_{\rm obs}$, $\delta$
(Eq. \ref{delta_vvb} below). Here the values of $g_{\rm N}$ for the observed
galaxies are computed from Eq. \ref{vcn}.  For most of the data points
of the simulated galaxies, the mass discrepancy is always larger than
for the observed galaxies. On the other hand, the observed data agree
extremely well with Milgromian dynamics (i.e., with there being no
cold or warm dark matter) with a standard $\mu -$function (orange
areas and symbols with error bars in Fig. \ref{disp_cdm}) and with a
simple $\mu$-function (red areas and symbols with error bars in
Fig. \ref{disp_cdm}).  The best fitting values of $a_0$ and the
corresponding errors with different $\mu$ functions are computed with
the Levenberg-Marquardt method, and are listed in the $6_{th}$ and
$7_{th}$ columns of Table \ref{sign}.  With a simple $\mu$-function
(red areas and symbols in Fig. \ref{disp_cdm}), the predictions of
mass discrepancy in Milgromian dynamics are slightly larger than the
observations. Therefore, the standard $\mu$-function is a better
interpolating function for the MDA correlations.

We also notice that the dispersion of the relation for the simulated
galaxies in the $\Lambda$CDM model is much wider than that of the
observational data, especially in the weak acceleration regime.
However, the correlation is tight for the observed galaxies. This is
surprising because the observational points have measurement
uncertainties which enlarge any spread. The wide spread of the
theoretical relation comes about because the spatial distribution of
dark matter does not tightly correlate with the baryonic distribution
in different simulated galaxies. Such a wide spread is due to the
multiplicity of free parameters of dark haloes: the parameters of the
dark matter profiles are not unique for different modeled disc
galaxies, the shapes of the dark haloes are triaxial, and there is a
variation between the inclination angle for the principle axes of the
dark haloes and the baryonic discs (as emphasized by
\citealt{Disney+2008}).  Instead, the observational data demonstrate a
very close one-to-one relationship between the baryons and their
rotation about the centre of their galaxy.

The lower panel of Fig. \ref{disp_cdm} shows the kinematic-luminous
acceleration relation of the simulated galaxies from $\Lambda$CDM
cosmological simulations (the coloured areas and shadows) and of the
observed galaxies (black points). The predictions from Milgromian
dynamic with two $\mu$-functions are plotted (red area for the simple
$\mu$-function and orange area for the standard $\mu$-function) as
well.  The kinematic-luminous acceleration relation of the simulated
galaxies lies above that of the observed galaxies, and there is a much
wider spread of such a relation for the simulated galaxies.  This is
not surprising, since the kinematic-luminous acceleration relation is
exactly the MDA correlation plotted differently and the MDA
correlation already indicates the disagreement between the simulated
and the observed galaxies.  Due to the stochastic galaxy formation
scenario in cosmological simulations \citep{Boylan-Kolchin+2011}, for
a given enclosed baryonic mass, there are various possibilities for
the centrifugal acceleration of the simulated galaxies. Therefore a
tight correlation is not expected from the model galaxies. For a
comparison, Milgromian dynamics predicts tight relations (with
different forms of the $\mu$-functions) for the MDA.

In summary, there are two problems for galaxy formation in
$\Lambda$CDM cosmology: (i) the fraction of CDM haloes with a
sufficiently quiet merging history is far too small to account for the
large fraction of galaxies that are bulgeless discs or disc-dominated
galaxies. (ii) Even those delicately selected DM haloes that do have a
quiet merging history fail to host disc galaxies which correspond to
real observed galaxies. Only the slow growth of the baryonic disc with
the DM halo without any mergers has yielded realistic-looking disc
galaxies, as already shown by \citet{Samland_Gerhard2003}. This is,
however, equivalent to a growing purely baryonic galaxy in Milgromian
dynamics.

\begin{figure}{}
\begin{center}
  \resizebox{9.cm}{!}{\includegraphics{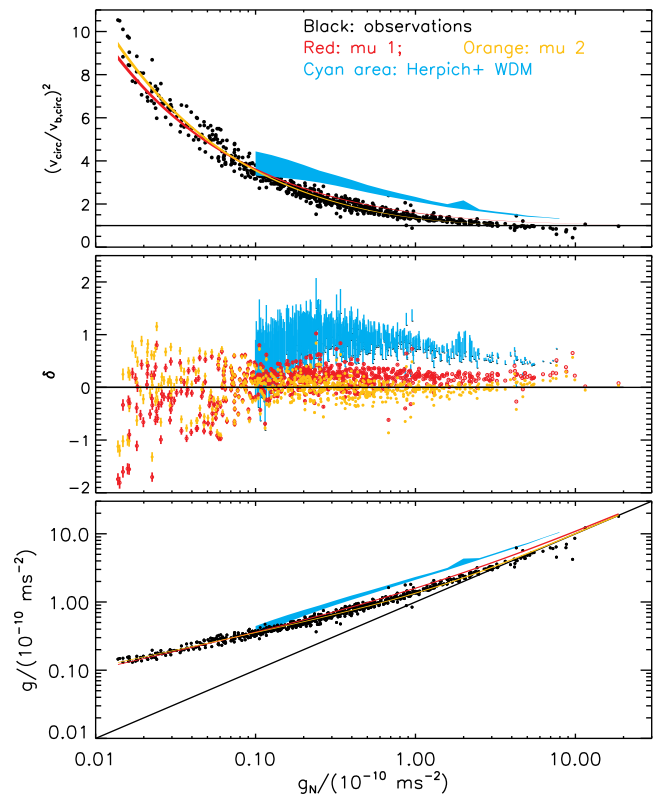}}
  \makeatletter\def\@captype{figure}\makeatother \caption{Upper panel:
    The black points and coloured areas are defined as in Fig.
    \ref{disp_cdm}. The cyan areas are the mass discrepancy-baryonic
    gravitational acceleration relation for simulated galaxies from
    warm dark matter \citep[][]{Herpich+2013} cosmological
    simulations.  Middle panel: the difference of the mass discrepancy
    between WDM models and observed galaxies, $\delta$
    (Eq. \ref{delta_vvb} below).  Lower panel: the kinematic-luminous
    acceleration relation of the CDM models and the observed galaxies
    (symbols and colours as in the upper panel). }\label{disp_wdm}
\end{center}
\end{figure}

\subsection{Disc galaxies from WDM cosmological simulations}\label{wdm}
\citet{Herpich+2013} simulated three galaxies, and the circular
velocities contributed from baryons and dark matter can be found for
two galaxies from the three (fig. 3 of Herpich et al. 2013), g1536 and
g5664, which are used here. For each galaxy, there are three sets of
parameters, corresponding to different masses of the WDM
particles. The MDA correlation for galaxies simulated in a WDM
cosmology is presented in the upper panel of Fig. \ref{disp_wdm} with
the cyan area. The difference of the MDA data between the WDM
simulated galaxies $\left[\frac{v(g_{\rm N})}{v_{\rm b}(g_{\rm N})}\right]^2_{\rm WDM}$
and the observations $\left[\frac{v(g_{\rm N})}{v_{\rm b}(g_{\rm N})}\right]^2_{\rm
  obs}$, $\delta$, is given by Eq. \ref{delta_vvb} below and is shown
in the middle panel. The kinematic-luminous-acceleration relation for
the WDM simulated galaxies is shown in the lower panel of
Fig. \ref{disp_wdm}. The MDA correlation is not consistent with the
observations. The relations obtained from the simulated galaxies lie
above the observational relation. Therefore, the above MDA data rule
out the two model galaxies obtained from WDM simulations as well.

In Fig. \ref{disp_wdm}, a comparison of the observational MDA data
predicted under the assumption of Milgromian dynamics with a standard
$\mu$-function (orange areas and symbols) and a simple $\mu -$function
(red areas and symbols) is shown. It confirms again that the
dispersion of the MDA data predicted by WDM models is larger than for
Milgromian dynamics with a standard $\mu$-function and a simple $\mu
-$function. This is a result of the spread of WDM halo properties for
a given baryonic mass (compare with \citealt{Disney+2008}).

\subsection{The Wilcoxon signed-rank test}\label{wilcoxon}
As mentioned above (Sec. \ref{cdm} and Sec. \ref{wdm}), the majority
of the MDA data obtained from the C/WDM simulated galaxies lie above
the relation obtained from the observations. Here we apply a
non-parametric statistical hypothesis test, the Wilcoxon signed-rank
test \citep{Wilcoxon1945,Bhattacharyya_Johnson1977}, to formally study
the difference of the data between the simulations and the
observations. The procedure of the Wilcoxon signed-rank test applied
here is as follows:
\begin{itemize}
\item 1. For the observed data (i.e., \citealt{McGaugh2004} data)
  there are $m$ data points. For each data point the values of $g_{\rm N}$
  (defined in Eq. \ref{vcn}) and the ratio
  $\left[\frac{v(g_{N})}{v_{\rm b}(g_{N})}\right]^2$ (data from
  \citealt{McGaugh2004}) are known. To compare the observed data with
  the simulated galaxies, interpolate the upper envelope of the MDA
  data of the simulated galaxies at each observed data point
  $g_{N,i}$, where $i=1,...,m$. Compute the difference of the MDA
  correlation between the simulated and the observed galaxies at each
  $g_{N,i}$, which is defined in Eq. \ref{vcn},

\beq \label{delta_vvb}
\delta_i=\left[\frac{v(g_{N,i})}{v_{\rm b}(g_{N,i})}\right]^2_{\rm
  C/WDM}-\left[\frac{v(g_{N,i})}{v_{\rm b}(g_{N,i})}\right]^2_{\rm obs}.
\eeq

\item 2. Exclude the simulated-observed data pairs with
  $|\delta_i|=0.0$. \footnote{This is the standard procedure of the
    Wilcoxon sign rank test. However, in the data sample used here,
    there is no data pair with $|\delta_i|=0.0$. The number of
    remaining data pairs is thus $m$.} Order the remaining data pairs
  according to increasing $|\delta_i|$. The number of the remaining
  data pairs is $n$. The rank of the data pairs is denoted as $R_i$.
\item 3. The data pairs with $\delta_i>0$ are selected, and the sum of
  their ranks are denoted as $T^+$. Let
  
\bey E&=&\frac{n(n+1)}{4},  \\
  D&=&\frac{n(n+1)(n+2)}{24}.  
\eey 

There could be a group of ties
  within which the $|\delta|$ values of the elements are the same,
  i.e., $|\delta_i|=|\delta_a|=|\delta_b|=...~(i\ne a\ne b \ne
  ...)$. Let 

\beq 
Q=\frac{1}{48} \sum_{j=1}^l q_j(q_j^2-1), 
\eeq 

where $l$ is the number of such ties, and $q_j$ is the number of
elements in the $j_{th}$ tie.
\item 4. Calculate the statistic $Z$ for a large sample of data pairs
  (for more details of the statistical method, see
  \citealt[][p. 519]{Bhattacharyya_Johnson1977}),

\beq Z=\frac{T^+-E}{\sqrt{D-Q}}.
\eeq

\item 5. Interpolate the lower envelope of the MDA data of the
  simulated galaxies at each $g_{N,i}$, and repeat the steps $1-4$.
\end{itemize}

To test how well the MDA correlation predicted by Milgromian dynamics
matches the observations, steps $1-4$ are repeated for Milgromian
dynamics with different $\mu$ functions. For each $\mu$ function, a
Levenberg-Marquardt fit to the observed MDA data is applied to obtain
the best fitting values of $a_0$ and the corresponding errors. The
results are listed in Table \ref{sign}. All of the CDM and WDM models,
together with Milgromian dynamics with the simple $\mu$-function, can
be ruled out with a confidence of better than $99.99\%$. Only under
Milgromian dynamics with the standard $\mu$-function, $Z=0.71$. Since
the level of significance for exclusion and for a directional
($1-$tailed) test is $\alpha=0.05$ for $Z=1.645$, Milgromian dynamics
with the standard $\mu$-function constitutes a good description of the
observed data. That is, the hypothesis that Milgromian dynamics/MOND
does not agree with the MDA data can be ruled out with at most
$1-2\alpha=90\%$ confidence.

Finally, we test the MDA correlation in the pure SID regime
(Sec.~\ref{STSI}).  Steps $1-4$ are repeated for the data pairs with
weak accelerations, $g_{\rm N}<0.2\times 10^{-10}\mssq$, and the best
fitting value of $a_0 =1.24\times 10^{-10}\mssq$
(Tab. \ref{sign}). $Z=-0.38$ is obtained.  This is an extremely good
agreement between SID and the observed data. The level of significance
for a $1-$tailed standard normal critical value is $-1.282$
corresponding to $\alpha=0.1$. Thus the hypothesis that SID does not
agree with the MDA data in the weak field regime can be rejected with
at most $1-2\alpha=80\%$ confidence.

\begin{table*}
\begin{center}\vskip 0.00cm
  \caption{Wilcoxon signed-rank test. The size of the sample of
    non-zero-differences of simulated-observed data pairs is $n=730$.
    The first column tabulates the data source from the simulated
    galaxies, the $2_{nd}$ and $4_{th}$ columns contain the number of
    data pairs with $|\delta|>0.0$ for the lower and upper envelopes
    of the simulated galaxies, respectively. The $3_{rd}$ and $5_{th}$
    columns list, respectively, the statistic $Z$ for the lower and
    upper envelopes of the simulated galaxies.  The $6_{th}$ column
    lists the values of $a_0$ for different $\mu$ functions. For
    Guedes+'s model, there is only one simulated galaxy (no lower
    envelope). The results assuming SID and the transition regime are
    valid, with the simple and standard $\mu$-functions, are listed in
    the $8_{th}$ and $9_{th}$ lines, respectively. The bottom line
    show the Wilcoxon signed-rank test for pure SID in the weak field
    regime (i.e., without the $\mu$ function), where $g_{\rm N}<0.2\times
    10^{-10}\mssq$. $n^-$ is the number of the data pairs with
    $|\delta|<0.0$.}
\begin{tabular}{llllllllc}
\hline
 & Lower  &envelopes  &  Upper & envelopes  & best-fit $a_0$\\
DM simulated galaxies &  $n^+ $ & $Z$ &$n^+$ & $Z$ & $10^{-10} m s^{-2}$\\
\hline
Marinacci+ & 467 &24.84 & 539 & 28.41 & \\
Agertz+ & 278 & 20.41 & 286 & 20.70 &\\
Aumer+ & 424 & 10.53 & 681 & 32.05  &\\
Guedes+& -  & -  & 654 & 29.01  &\\
Herpich+ (WDM) & 527 & 28.27 & 536 & 28.33 & \\
MOND simple $\mu$ &- &-  & 574 & 18.86 & $0.94\pm 0.03$ \\
MOND standard $\mu$ & -&-& 375 & 0.71 & $1.21\pm 0.03$ \\
\hline
deep MOND & n& $n^-$ & $n^+$ & $Z$  & $a_0$\\
($g_{\rm N}<0.2 \times 10^{-10} \mssq $)  & 289 &154 & 135 & -0.38 & $1.24 \pm 0.03$\\
\hline
\end{tabular}
\label{sign}
\end{center}
\end{table*}

\section{The dark matter halo to stellar mass relation}\label{masses}
\subsection{The masses of CDM haloes}\label{CDMmasses}
\begin{table*}
\begin{center}\vskip 0.00cm
\caption{The right ascensions and declinations (epoch J2000, the $2_{\rm nd}$ and
$3_{\rm rd}$ columns), stellar masses (the $4_{\rm th}$ column) and
halo masses (the $5_{\rm th}$ column) of dwarf galaxies in
\citet{Miller+2013}. The strength of the external field 
(Eq. \ref{ge}) is listed in
the $6_{\rm th}$ column for the labeled magenta dwarf galaxies in
Fig. \ref{tf}, assuming $M_{\rm b}=M_s$. Here $a_0=1.21\times10^{-10} \mssq $ as in Tab. \ref{sign}.}
\begin{tabular}{llllllllc}
\hline
Number & R.A. & Dec. & Log$(M_s/\msun)$ & Log$(M_{\rm vir}/\msun)$ & $g_e/a_0$ \\
\hline
1  & 34.287564 & -5.1435433 & 8.74 $\pm$ 0.09 & 10.79 $\pm$ 0.03 & \\
2  & 34.303787 & -5.1458641 & 8.32 $\pm$ 0.12 & 10.39 $\pm$ 0.15 & \\
3  & 34.358524 & -5.1515666 & 7.84 $\pm$ 0.09 & 9.36 $\pm$ 0.39  & \\
4  & 34.398975 & -5.1525832 & 8.79 $\pm$ 0.07 & 10.70 $\pm$ 0.14 & \\
5  & 34.289186 & -5.1563347 & 8.31 $\pm$ 0.09 & 9.80 $\pm$ 0.35  & \\
6  & 34.359126 & -5.1571952 & 7.29 $\pm$ 0.14 & 9.68 $\pm$ 0.30  & \\
7  & 34.336092 & -5.1585409 & 7.83 $\pm$ 0.08 & 9.56 $\pm$ 0.13  & 0.019\\
8  & 34.489719 & -5.1596155 & 8.64 $\pm$ 0.14 & 10.78 $\pm$ 0.32 & \\
9  & 34.404135 & -5.1638168 & 8.43 $\pm$ 0.06 & 10.20 $\pm$ 0.09 & \\
10 & 34.310380 & -5.1642154 & 8.54 $\pm$ 0.13 & 10.44 $\pm$ 0.20 & \\
11 & 34.367527 & -5.1671357 & 8.41 $\pm$ 0.08 & 10.47 $\pm$ 0.27 & \\
12 & 34.440428 & -5.1676006 & 8.74 $\pm$ 0.09 & 10.55 $\pm$ 0.21 & \\
13 & 34.473897 & -5.1691046 & 7.76 $\pm$ 0.11 & 9.71 $\pm$ 0.25  & \\
14 & 34.446086 & -5.1700983 & 8.15 $\pm$ 0.15 & 9.71 $\pm$ 0.08  & 0.028\\
15 & 34.279085 & -5.1710765 & 8.30 $\pm$ 0.06 & 10.38 $\pm$ 0.16 & \\
16 & 34.350739 & -5.1712933 & 7.25 $\pm$ 0.11 & 9.26 $\pm$ 0.18  & \\
17 & 34.503690 & -5.1738860 & 8.27 $\pm$ 0.08 & 10.09 $\pm$ 0.18 & \\
18 & 34.432528 & -5.1760097 & 8.75 $\pm$ 0.09 & 10.64 $\pm$ 0.06 & \\
19 & 34.466202 & -5.1780329 & 8.76 $\pm$ 0.10 & 10.46 $\pm$ 0.03 & 0.020\\
20 & 34.429762 & -5.1785330 & 8.96 $\pm$ 0.06 & 11.09 $\pm$ 0.17 & \\
21 & 34.434409 & -5.1793119 & 8.50 $\pm$ 0.08 & 10.46 $\pm$ 0.40 & \\
22 & 34.517196 & -5.1793529 & 7.89 $\pm$ 0.07 & 9.68 $\pm$ 0.20  & \\
23 & 34.250846 & -5.1797838 & 7.60 $\pm$ 0.13 & 9.92 $\pm$ 0.09  & \\
24 & 34.415249 & -5.1820298 & 8.50 $\pm$ 0.08 & 10.39 $\pm$ 0.06 & \\
25 & 34.296140 & -5.1827143 & 8.13 $\pm$ 0.12 & 9.94 $\pm$ 0.06  & 0.015\\
26 & 34.362769 & -5.1829861 & 8.03 $\pm$ 0.09 & 9.67 $\pm$ 0.08  & 0.023\\
27 & 34.477748 & -5.1834746 & 7.46 $\pm$ 0.11 & 9.75 $\pm$ 0.03  & \\
28 & 34.423177 & -5.1861241 & 7.95 $\pm$ 0.08 & 9.92 $\pm$ 0.23  & \\
29 & 34.282862 & -5.1864030 & 8.02 $\pm$ 0.07 & 10.11 $\pm$ 0.15 & \\
30 & 34.520764 & -5.1871599 & 7.06 $\pm$ 0.27 & 9.31 $\pm$ 0.31  & \\
31 & 34.413980 & -5.1882020 & 8.20 $\pm$ 0.20 & 9.96 $\pm$ 0.09  & 0.017\\
32 & 34.523446 & -5.1899136 & 7.40 $\pm$ 0.21 & 9.62 $\pm$ 0.21  & \\
33 & 34.254774 & -5.1914894 & 8.50 $\pm$ 0.08 & 10.45 $\pm$ 0.19 & \\
34 & 34.359852 & -5.1916576 & 8.66 $\pm$ 0.14 & 10.30 $\pm$ 0.35 & \\
35 & 34.417074 & -5.1950333 & 8.36 $\pm$ 0.11 & 9.95 $\pm$ 0.13  & 0.026\\
36 & 34.370424 & -5.2035566 & 8.80 $\pm$ 0.11 & 10.67 $\pm$ 0.02 & 0.013\\
37 & 34.465397 & -5.2073517 & 8.82 $\pm$ 0.07 & 10.64 $\pm$ 0.13 & \\
38 & 34.446346 & -5.1498754 & 9.24 $\pm$ 0.12 & 10.95 $\pm$ 0.09 & \\
39 & 34.283345 & -5.1519078 & 9.06 $\pm$ 0.13 & 10.96 $\pm$ 0.08 & \\
40 & 34.392003 & -5.1590678 & 8.94 $\pm$ 0.07 & 10.83 $\pm$ 0.17 & \\
41 & 34.275351 & -5.1724717 & 9.08 $\pm$ 0.16 & 10.96 $\pm$ 0.05 & \\
\hline
\end{tabular}
\label{s2h}
\end{center}
\end{table*}

Above we have seen that the most advanced models based on
Einsteinian/Newtonian gravitation together with cold or warm dark
matter (DM) are not able to reproduce the observed MDA
correlation. Another related way to test for the existence of dark
matter haloes is to study the dark-matter-halo-mass versus
baryonic-mass correlation. Recently, \citet{Miller+2013} constrained
the stellar to halo mass relation for a sample of 41 dwarf galaxies
within a redshift range of $0 < z < 1$ (the positions, stellar and
dark halo masses of the galaxies are listed in Tab. \ref{s2h}): the
stellar masses, $M_s$, are derived from the Fitting and Assessment of
Synthetic Templates \citep{Kriek+2009} code for the photometric
database \citep{Newman+2013}, and the DM halo masses, $\mvir$, are
computed from 

\beq \label{mv200} {\rm log}_{10} (\mvir /h^{-1}) = 3 \, {\rm log}_{10}
(G^{-1}V_{200} ) 
\eeq

\noindent
assuming the haloes are spherical. Here $h=H_0/100\kms\mpc^{-1}$,
$V_{200}$ is the rotation speed at the virial radius, $\rvir$, within
which the enclosed CDM halo mass has a mean overdensity of 200 times
of the critical density of the Universe, $\rho_{\rm
  crit}=\frac{3H_0^2}{8\pi G}$.  The parameters of their cosmological
model are $\Omega_\Lambda=0.7,~\Omega_m=0.3,~H_0=70 \kms \mpc^{-1}$.
$V_{200}$ is converted according to

\beq\label{v22v200}
V_{2.2}/V_{200}=1.05,
\eeq 

\noindent where $V_{2.2}$ is a direct measurement of the circular
velocity at the radius $r_{2.2}$ which is 2.2 times the scale radius
(for more details, see Sec. 3.3 and Sec.4 in \citealt{Miller+2013}).
Note that this is an empirical relation measured from weak lensing by \citet{Reyes+2012} for galaxies within the mass range of $[10^9, ~10^{11}]\msun$, and the slope of velocity ratio to stellar masses is $0.53\pm0.03$. \citet{Miller+2013} extrapolated this relation to low mass dwarf galaxies with masses of $10^7\msun$, thus the dispersion of $V_{200}$ of the haloes derived from this relation
is largest for the low-mass
galaxies (see figures 6-7 in Miller et al. 2014). Since the 'virial
mass' in \cite{Miller+2013} is not derived from dynamics, it does not
exactly amount to the virial mass defined in Eq. \ref{ms2mh_cdm} and
Eq. \ref{mvir} below.  

\citet{Miller+2013} compared their observations with CDM cosmological
simulations by \citet{Behroozi+2013} within a similar redshift
range. They found that the stellar to halo mass relation predicted
from the simulated dwarf galaxies is at odds with observations. For a
given stellar mass, the simulations significantly over-predict the
mass of dark matter for the dwarf galaxies.  Although the dispersion
of data for the dwarf galaxies is large, the trend is clear that the
observed data are not consistent with the curves from the simulated
galaxies.  This problem is essentially the same as that of the MDA
correlation at large galactic radii, i.e., at low acceleration (see
Fig. \ref{disp_cdm}-\ref{disp_wdm}).

Furthermore, \citet{Guo+2010} proposed an abundance-matching stellar
to halo mass relation for model galaxies from $\Lambda$CDM
cosmological simulations. However, in the low stellar mass range
$[10^6,10^8]\msun$, the theoretical halo masses are too large by a
factor of 5 compared to those of dwarf galaxies in a large survey of
SDSS central galaxies\footnote{In \citet{Guo+2010}, the CDM
  (sub-)haloes are selected from all (sub-)haloes that have galaxies at
  their centres.} \citep{More+2009}. \citet{Ferrero+2012} improved the
\citet{Guo+2010} stellar to halo mass relation as follows:

\beq\label{ms2mh_cdm} 
\frac{M_s}{M_{\rm
    vir}}=c\left[1+\left(\frac{M_{\rm
        vir}}{M_1}\right)^{-2}\right]^\kappa\left[\left(\frac{M_{\rm
        vir}}{M_0}\right)^{-\alpha}+\left(\frac{M_{\rm
        vir}}{M_0}\right)^{\beta}\right]^{-\gamma}, 
\eeq 

\noindent where $c=0.129$, $M_0=10^{11.4}\msun$,
$M_1=10^{10.65}\msun$, $\alpha=0.926$, $\beta=0.261$ and
$\gamma=2.440$. $\kappa$ is a free parameter.  Larger values of
$\kappa$ represent shallower stellar to halo mass relations for low
mass haloes, while $\kappa=0$ returns to the abundance-matching
relation proposed by \citet{Guo+2010}. We note that the procedure to
associate visible galaxies with their dark matter haloes which are
derived from dark-matter-only simulations lacks the physics of galaxy
formation entirely. Abundance-matching is merely the short-circuiting
of a major problem of $\Lambda$CDM cosmology (the ``missing dwarf
galaxy problem'', or more truthfully ``the dwarf over-prediction
problem'').

\begin{figure*}{}
\begin{center}
  \resizebox{11.5cm}{!}{\includegraphics{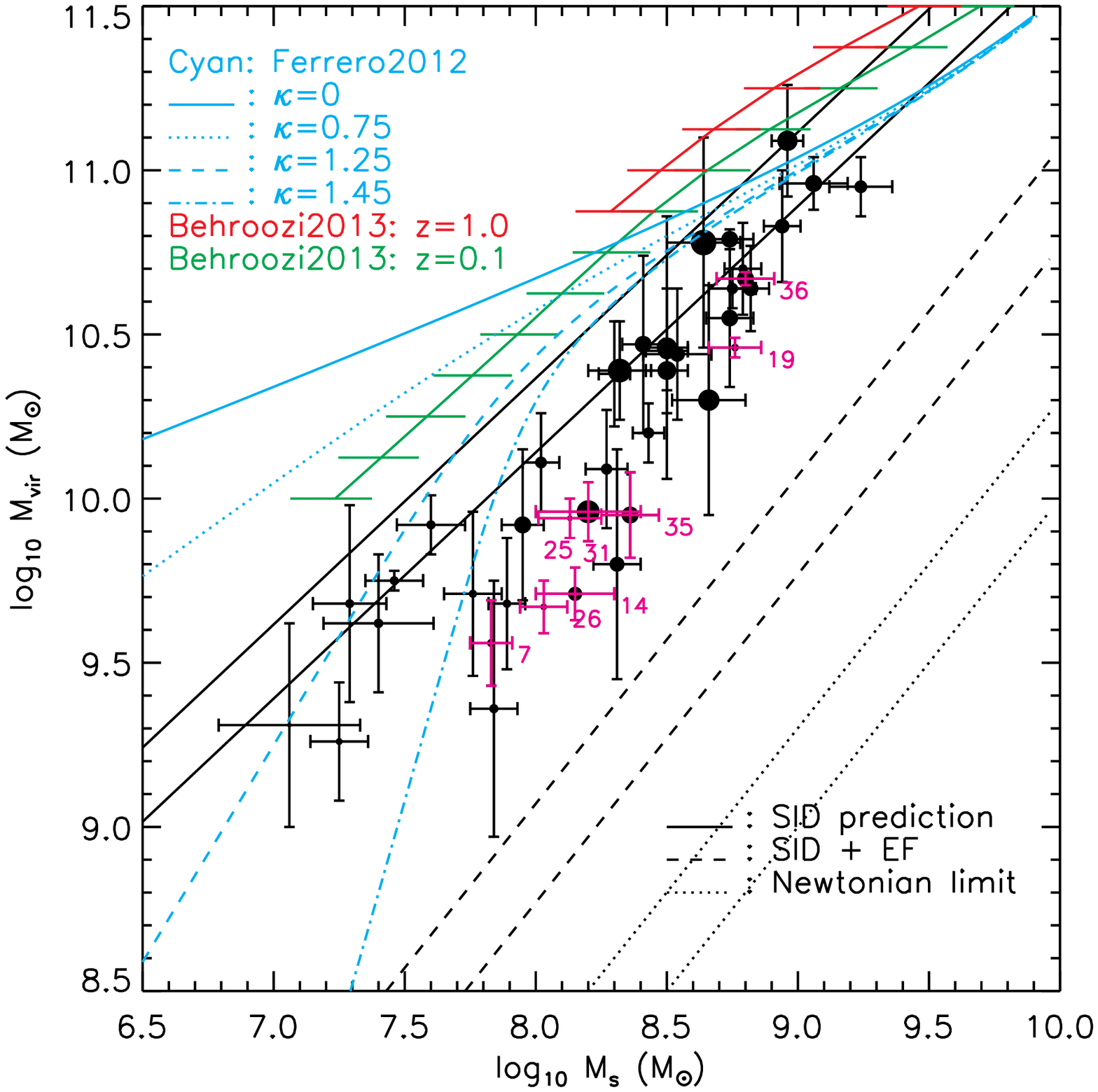}}
  \makeatletter\def\@captype{figure}\makeatother \caption{The halo to
    baryonic (stellar plus gas) mass relation from observations by
    \citet[][black circles with error bars, the size of the symbols
    represents the redshift, $z\in(0,1)$, of the dwarf galaxies:
    larger symbols for higher redshifts and smaller symbols for lower
    redshifts]{Miller+2013} and from simulated galaxies by
    \citet[][cyan curves]{Ferrero+2012} and \citet[][red and green
    curves, corresponding to $z=1.0$ and $0.1$,
    respectively]{Behroozi+2013}. The black lines are predictions from
    SID (Eq.~\ref{mh2mb}): for isolated galaxies (solid lines, the
    upper is for assumming the mass of gaseous matter, $M_g$, equals
    to the mass of stars, $M_s$, in a galaxy and the lower solid line
    is for assumming the mass of gas $M_g=0$ in a galaxy), for
    galaxies located at the position of the LMC near a Milky-Way-like
    galaxy (dashed lines, the upper and lower dashed lines are defined
    the same as solid lines) and for galaxies embedded in a strong
    external field (Newtonian limit, dotted lines, the upper and lower
    dashed lines are defined the same as solid lines).  The magenta
    symbols point out the dwarf galaxies whose halo-to-baryonic mass
    ratio lies beyond the $3\sigma$ confidence level away from the
    prediction of SID, i.e., galaxies probably embedded in external
    fields. The magenta numbers show the numbers of the corresponding
    galaxies in Table \ref{s2h}. 
    }\label{tf}
\end{center}
\end{figure*}

We compare the halo to stellar masses predicted by Eq.
\ref{ms2mh_cdm} (cyan curves) for different values of $\kappa$ to
the empirical data of \citet{Miller+2013} in Fig. \ref{tf}, finding
an inconsistency between the simulated and observed galaxies. For a
given stellar mass, the halo mass is significantly over-predicted for Ferrero's
relation for all different values of $\kappa$.

\subsection{The mass of the phantom dark matter halo}\label{pdmmass}

Concerning Milgromian dynamics, a baryonic object is surrounded by an
unreal non-particle (i.e., phantom) isothermal dark matter halo with
constant circular velocity given by Eq. \ref{vc}. This follows
  directly from pure-SID, i.e. even without considering the transition
  from pure-SID to the Newtonian regime, the description of which
  constitutes Milgromian dynamics (Appendix~\ref{MOND}) in the
  classical dynamical regime. The apparent phantom virial dark
matter halo mass, $M_{\rm vir}$, can be derived from Eq. \ref{vc}
\citep[for the derivation, see Sec. 2.1 in][]{Wu_Kroupa2013}, which is
\footnote{The virial mass in SID can be derived in the same way as in
  \citet{Miller+2013}, i.e., from a relation between $V_{200}$ and
  $V_{2.2}$ in SID. The derivation of such a ``virial mass'' is
  presented in Appendix~\ref{mess}. However, because the errors of the
  Tully-Fisher (TF) relation in the galaxies of \cite{Miller+2013} are
  large, and the fit parameters for the TF relation are significantly
  different for different samples of galaxies (see table~2 in
  \citealt{Miller+2013}), the so-obtained ``virial mass'' is strongly
  sample dependent. Therefore a more universal virial mass of a galaxy
  in SID is adopted in Sec.~\ref{pdmmass}. Essentially, we assume that
  the inner rotation velocity (see Sec.~\ref{CDMmasses}) is a measure
  of the flat part of the rotation velocity at larger radii, which in
  turn measures the mass of the dark matter halo.}

\bey\label{mvir} M_{\rm vir}&=&(Ga_0M_{\rm b})^{3/4} p^{-1/2}G^{-3/2}, \\
p&=&\frac{4}{3}\pi \times 200 \rho_{\rm crit}.  \nonumber \eey 

\noindent Such phantom dark matter haloes are associated with a
non-inertial (i.e., unreal) mass by a Newtonian observer. {\it That
  is, an observer interpreting the motion of a star around a galaxy in
  terms of Newtonian dynamics will deduce (wrongly) that the galaxy is
  immersed in a dark matter halo. However, the inertial mass of the
  galaxy is exactly its baryonic mass only.}

Therefore, the phantom dark matter halo mass is sourced entirely by
the baryonic mass of the galaxy (Eq.~\ref{mvir}).  The phantom halo to
baryonic (stellar plus gas) mass ratio follows from Eq.~\ref{mvir},

\beq\label{mh2mb} \frac{M_{\rm
    vir}}{M_{\rm b}}=(Ga_0)^{3/4} M_{\rm b}^{-1/4}p^{-1/2}G^{-3/2}, 
\eeq 

\noindent which is shown with two black solid lines in Fig. \ref{tf}.
Since the amount of gas is difficult to determine in the observations,
the mass of gas is a parameter in SID/MOND which is not well
constrained. Here $a_0=1.21\times 10^{-10} ms^{-2}$, which is the
value determined by \citet{Begeman+1991}, and agrees with the best-fit
value of the MDA data in Sec. \ref{cdm}.  The upper limit for the mass
of gas, $M_g$, is assumed to be the same as $M_s$, i.e.,
$M_{\rm b}=2M_s$. The lower limit for the mass of gas is $M_g=0$, i.e.,
$M_{\rm b}=M_s$. The two limits correspond to the upper and lower solid
lines in Fig. \ref{tf}. These two assumptions on the mass of gas most
probably straddle the real gas content of the dwarf galaxies which is
undetermined for the sample of \citet{Miller+2013}.  The existence of
gas decreases the halo to stellar mass ratio in Milgromian
dynamics. There are overlaps between the prediction of Milgromian
dynamics and CDM cosmological simulations for the isolated gas-rich
galaxies ($M_g=M_s$).  However, the prediction based on Milgromian
dynamics agrees much better with the observations within the error
range of the data. The data which deviate more than $3\sigma$ from
Eq. \ref{mh2mb} are over-plotted using magenta symbols in
Fig. \ref{tf}. These data points are possibly the bad points of the
sample, since the virial mass for the data points are obtained from
the empirical relation of Eq. \ref{v22v200}, and they are not exactly
the virial mass derived from dark halo dynamics. Interestingly, SID
predicts a truncated phantom halo which stays below the solid lines in
the figure if its baryonic source system is exposed to an external
field (see Sec. \ref{efe}).

\subsection{The external field effect in SID}\label{efe}

\subsubsection{The virial mass of a galaxy embedded in an external field}\label{virial}

In SID/Milgromian dynamics there appears an interesting effect that
leads to an observable prediction which does not exist in the
Newtonian plus dark matter model. It is relevant for a satellite
object falling within the gravitational field of a host such that the
host field does not vary significantly across the satellite. This
external gravitational field effectively truncates the isothermal
phantom dark matter halo, as deduced by an observer who interprets the
observations within Newtonian dynamics. The Milgromian dynamics/MOND
equation for a spherical, axisymmetric or cylindrical system embedded
in an external field is \citep{Milgrom1983a,Sanders_McGaugh2002}
\beq\label{efinterpol}
{\bf  g}_{N,i}+{\bf g}_{N,e}=\mu(|{\bf (g_i + g_e)}|/a_0) ({\bf g_i}+{\bf g_e}),
\eeq 
\noindent where ${\bf g}_{N,i}$ is the Newtonian acceleration from the baryonic matter of the internal system, ${\bf g}_{N,e}$ is the Newtonian acceleration from the baryonic matter of the external gravitational source generating the background uniform field, and ${\bf g_i}$ and ${\bf g_e}$ are the internal and external gravitational accelerations. For an external field dominated system, $g_e=|{\bf g_e}| \gg g_i=|{\bf g_i}|$, Eq. \ref{efinterpol} is expanded around $g_e$ to lowest order as 
\bey\label{efzeroth}
g_{N,e}=\mu(|{\bf g_e}|/a_0) {\bf g_e},\nonumber \\
g_{N,i}=\mu(|{\bf g_e}|/a_0) {\bf g_i},
\eey
with a dilation factor of $\Delta_1=(1+d\ln \mu/d \ln x)_{x=g_e/a_0}$. $\Delta_1$ approaches $1$ in the Newtonian limit and approaches $2$ in the deep MOND limit. The value of $\Delta_1$ also depends on the direction relative to the external field (for more details see \citealt{Milgrom1983a,BM1984,Zhao_Tian2006}).
$\mu(x)=x$ in the low acceleration regime where SID is valid (i.e., in deep MOND limit).\footnote{Note that SID as such does NOT necessary imply the introduction of $a_0$, nor the role of acceleration. One can obtain SID by introducing a length scale factor, $r_0$, instead of the introduction of $a_0$, into the Newtonian law of gravity. Thus $g=GM/(rr_0)$ instead of the Newtonian spherical symmetric gravity, $g_{\rm N}=GM/r^2$. One can also obtain SID by introducing a time constant, $t_0$, such that the gravity becomes $g=(GM/t_0)^{2/3}r^{-1}$. Therefore SID is obtained from the above two examples, but the baryonic Tully-Fisher relations are different. $M\propto v^2$ for the length scale factor in the former case and $M\propto v^3$ for the time constant in the latter case, whereas $M\propto v^4$ through the introduction of $a_0$ in SID/MOND. Therefore, MOND is based on two main axioms: the introduction of $a_0$ as the role of an acceleration constant and SID is the deep MOND limit.}
 
A star on a circular orbit at a large distance from the satellite
galaxy with mass $M_{\rm b}$ orbits subject to the centrifugal acceleration
$g_i=v^2/r$. In SID \citep{Milgrom2009a,Milgrom2014b} a
Newtonian observer interprets this to be due to the centripetal
acceleration from an isothermal (phantom) dark matter halo (PDMH) with
mass within radius $r$ of $M_{\rm PDMH}(<r)$. Assuming spherical
symmetry, it follows that

\beq\label{dm_mond}
\frac{GM_{\rm PDMH}(<r)}{r^2} = \frac{\sqrt{GM_{\rm b}a_0}}{r}.
\eeq

\noindent Thus, the mass of the phantom dark matter halo is 

\beq\label{mpdmh} 
M_{\rm PDMH}(<r) =(M_{\rm b}a_0)^{1/2}G^{-1/2}r. 
\eeq

\noindent An isolated galaxy has an infinitely extended PDMH with an
unbounded phantom mass.  An isolated galaxy within a cosmological
model has a virial PDMH mass given by Eq. \ref{mvir} above which
follows from equating the mean PDMH density within $r_{\rm vir}$ to
200 times the critical density in the universe, yielding the maximum
radius $r_{\rm vir}$ of the isothermal PDMH. If the galaxy is immersed
in a uniform external gravitational field corresponding to an
acceleration ${\bf g}_e=(0,~0,~g_e)$ on a Cartesion grid, i.e., the
external field is along the $z -$axis direction, then the centripetal
acceleration from the PDMH, $g_i$, equals $g_e=|{\bf g}_e|$ at the
radius $r_{\rm eq, SID}=v^2/g_e =\sqrt{GM_{\rm b}a_0}/g_e$. Using
Eq. \ref{mpdmh}, the PDMH mass of such a galaxy is thus reduced in
mass to the value

\begin{equation}
M_{\rm PDMH}(r_{\rm eq, SID}) = (M_{\rm b}a_0)^{1/2}G^{-1/2} r_{\rm eq,
SID}. \label{pdm}
\end{equation}

\noindent
At radius $r>r_{\rm eq, SID}$ the star accelerates mainly according to
${\bf g}_e$, while at $r<r_{\rm eq, SID}$ it accelerates mainly
according to internal ${\bf g}_i$.

Thus, a strict prediction following from SID and thus from Milgromian
dynamics (i.e. MOND) is that a Newtonian observer will deduce galaxies
to have a maximal (phantom) dark matter halo mass given by $M_{\rm
  PDMH}=M_{\rm vir}$ (Eq. \ref{mvir}). Galaxies which are immersed in
a uniform external field will appear to have reduced PDMH masses
(Eq. \ref{pdm}).  The two solid lines in Fig. \ref{tf} show the
virial masses of isolated galaxies in SID. SID  predicts that
galaxies embedded in external fields stay below the solid lines in
Fig. \ref{tf}.

\subsubsection{A distance---strength-of-external-field relation}

At $r=r_{\rm eq, SID}$, the strength of the external field as obtained
by a Newtonian observer is 

\beq\label{ge_r}
g_e=\frac{GM_{\rm PDMH}(r_{\rm eq, SID})}{r_{\rm eq, SID}^{2}}. 
\eeq

\noindent
For a given dwarf galaxy, the strength of the external field as a
function of the observationally determined baryonic mass and the
observationally determined phantom dark matter halo mass can be
derived from the combination of $g_i(r_{\rm eq, SID})=g_e$,
Eq. \ref{vc} and Eq.  \ref{ge_r},

\beq\label{ge}
g_e=\frac{M_{\rm b}}{M_{PDMH}}a_0, 
\eeq

\noindent 
where $M_{PDMH}$ is the unreal phantom dark matter halo mass of the
dwarf galaxy observationally deduced by a Newtonian observer. This
simple relation supplies a quick way to determine the strength of an
external field which a dwarf galaxy is exposed to if the baryonic and
halo (or dynamical) mass of the galaxy are known.

For galaxies whose phantom halo masses deviate from the SID prediction
(Eq.~\ref{mh2mb}) by more than the $3\sigma$ confidence level (the
magenta symbols and numbered labels in Fig.  \ref{tf}), the strength
of their external fields (Eq. \ref{ge}) are computed and listed in
Table \ref{s2h}. The external fields are generally weak, from
$0.01a_0$ to $0.03a_0$. In SID/Milgromian dynamics all the data points
are expected to stay on (for the field galaxies) and below (for the
dwarf galaxies near another gravitational source) the solid lines in
Fig. \ref{tf}: upper solid line for gas-rich galaxies and lower
solid line for gas-poor galaxies.

Eq. \ref{ge} is a relation between the baryonic mass, $M_{b,\rm
  host}$, of a nearby galaxy or cluster of galaxies generating the
required external field and the distance to the centre of the nearby
gravitational source, $d$. For a given $g_e$, for example, as
calculated from Eq. \ref{ge}, $d=v_{\rm host}^2/g_e$, where $v_{\rm
  host}=(GM_{b,\rm host}a_0)^{1/4}$, thus 

\beq\label{md}
d=\sqrt{GM_{b,\rm host}a_0}/g_e.
\eeq 

\begin{figure}{}
\begin{center}
  \resizebox{9.cm}{!}{\includegraphics{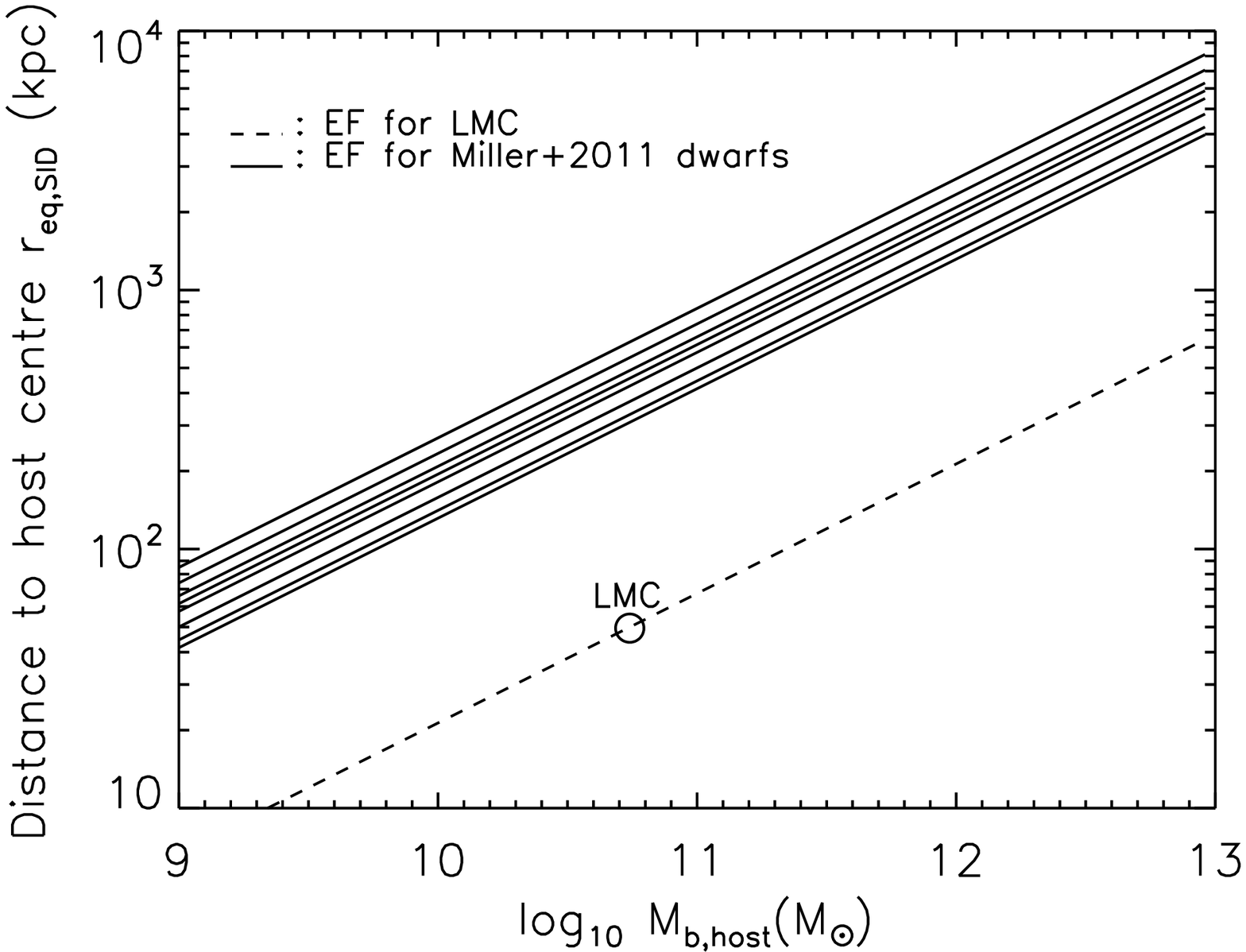}}
  \makeatletter\def\@captype{figure}\makeatother \caption{The
    $d(M_{b,\rm host})$ relation (Eq. \ref{md}) for the dwarf galaxies
    of \citet{Miller+2013} assuming they are embedded in an external
    field. Here $d$ is the distance between a dwarf galaxy in the
    \citet{Miller+2013} sample (labeled magenta in Fig. \ref{tf}, from
    top to bottom are for models 36, 25, 31, 7, 18, 26, 35, 14 in
    Tab. \ref{tf}) and the centre of the external field source (solid
    lines). $M_{b,\rm host}$ is the baryonic mass of the external
    field souce, which could be a nearby galaxy or a cluster of
    galaxies. The dashed line is the $d(M_{b,\rm host})$ relation
    within an external field of $0.17a_0$, which is the strength of
    the external field at the LMC due to the Milky Way. The circle is
    the ($d,~M_{b,\rm host}$) position of the LMC.  }\label{host}
\end{center}
\end{figure}

\noindent 
Here $v_{\rm host}$ is the circular velocity of the host-galaxy's
phantom dark matter halo generated by the host galaxy's baryonic mass,
$M_{b,\rm host}$.  Fig. \ref{host} shows the $d(M_{b,\rm host})$
relation for the dwarf galaxies of \citet{Miller+2013} assuming they
are embedded in external fields, i.e., for the magenta symbols in
Fig. \ref{tf}, and assuming that for the dwarfs $M_{\rm b}=M_s, M_g=0$. If
the nearby galaxy is also a dwarf galaxy, with a baryonic mass of
$10^9\msun$, the distances to the dwarf galaxies are only $40-90\kpc$,
while if the nearby external field source is a rich cluster of
galaxies with a baryonic mass of $10^{13}\msun$, the distances between
the cluster centre and the Miller et al. (2013) dwarf galaxies are
$4000-9000\kpc$.

For a comparison, we consider the Large Magellanic Cloud (LMC, dashed
lines in Fig. \ref{tf}, the upper line is for gas-rich galaxies,
i.e., with the assumption of $M_{\rm b}=2M_s, M_g=M_s$, and the lower line
is for gas-poor galaxies, i.e. with the assumption
$M_{\rm b}=M_s, M_g=0$), which is located at a Galactocentric distance of
$49.5~\kpc$ \citep{Kallivayalil+2006}. The LMC is embedded in the
gravitational background field of the Milky Way (MW). To calculate the
external field strength of the MW at the position of the LMC, the
baryonic Milky Way model with an interpolating function of `simple'
$\mu$ form is used from \citet{Wu+2008}. For the MW as a host galaxy,
$g_e\approx 0.17a_0$ for the LMC can be obtained from this model. All
the 41 galaxies in Fig. \ref{tf} are above the halo to stellar mass
relation of the LMC, since they are either isolated galaxies or
embedded in external fields (see the $6_{\rm th}$ column of
Tab. \ref{s2h}) in general one order of magnitude smaller than that of
the LMC. As a result, the phantom dark matter haloes have larger
truncation radii, and the enclosed halo masses within the truncation
radii are larger. The dashed line in Fig. \ref{host} shows the
$d(M_{b,\rm host})$ relation for the external field of the LMC, and
the circle represents the values of $d$ and $M_{b,\rm host}$ for the
LMC in the Milky Way. For the dwarf galaxies of \citet{Miller+2013}
within an external field, i.e., the magenta symbols in Fig. \ref{tf},
if the nearby galaxy is a Milky-Way like galaxy, the distances between
the galaxy sourcing the external field and the dwarfs are about
$200-400\kpc$ (see Fig. \ref{host}).

SID/Milgromian dynamics predicts that for dwarf galaxies, the halo to
baryonic mass relation has to stay in the range between the upper
solid line (isolated galaxies) and the lower dashed line (galaxies
embedded in an external field from a nearby massive galaxy) in
Fig. \ref{tf}. {\it Thus it is possible to falsify MOND by future
  observations of the stellar to halo mass relation of dwarf
  galaxies.} For example, an isolated field dwarf galaxy must appear
near the solid lines of Fig. \ref{tf} since MOND would be falsified
otherwise. Also, a self-gravitation satellite star cluster or dwarf
galaxy will, as a result of the external field, compress as it orbits
away from its host due to the build-up of its phantom dark matter
halo, and conversely, it will expand on its orbit towards its host
\citep{Wu_Kroupa2013}. The same satellite will be larger in size near
a more massive host. This may be the explanation why the Andromeda
satellites have larger radii compared to the satellite galaxies of the
Milky Way \citep{Collins+2011}.

\section{Conclusions}\label{conc}

The observed centrifugal acceleration and the Newtonian acceleration
for disc galaxies are extremely strongly correlated. In the weak field
regime, where $g_{\rm N} << a_0$, the correlation follows readily from SID
(Sec.~\ref{STSI}).

We showed that the mass discrepancy and the acceleration are
correlated for disc galaxies simulated in CDM and WDM cosmological
models. But, the correlation does not agree with the observations.
This indicates that the best simulated disc galaxies, which are
delicately chosen to have had a quiet merger history, even so still do not
represent the correct rotation curves. In any case, the fraction of
real galaxies without a bulge is far larger than the fraction of model
galaxies in the dark matter framework which have no significant merger
history, thus making the dark matter framework unlikely to work. The here reported analysis thus adds to the growing evidence that cold or warm dark matter does not exist. If so, then dynamical friction between galaxies on their dark matter halos would not be evident in the galaxy population as galaxies would merge rarely \citep{Kroupa2014}. Recent additional evidence for a lack of mergers has been independently found by \citet{Lena+2014}.

Moreover, the dispersion of the MDA correlation in the weak
acceleration regime that is obtained in the CDM and the WDM
simulations is much wider than for the observational data. The tight
correlation between the observed centrifugal acceleration and the
Newtonian acceleration from the observed baryon-only masses implies a
small scatter of the baryonic Tully-Fisher relation
\citep{McGaugh2004}, which has been confirmed by the more recent
observations of \citet{McGaugh2012}. On the other hand, the wide
spread of the relation in CDM and WDM simulations indicates a large
scatter of the baryonic Tully-Fisher relation, which is incompatible
with the observations. The large scatter in the simulated $\Lambda$CDM
and the WDM galaxies is a necessary feature of these models because
the dark matter haloes have various shapes and masses for a given
baryonic galaxy, as already noted by \citet{Disney+2008}.  Instead,
the Newtonian observer detects unphysical (i.e., phantom) dark matter
halo masses which are due to Milgromian rather than Newtonian
dynamics in the weak field regime.

The halo to baryonic mass relation is studied for cosmologically
simulated galaxies and for SID. We find that SID (and thus MOND)
predicts an apparent (phantom) virial halo to baryonic mass relation
which agrees well with observations. The simulated galaxies from CDM
models fail to reproduce the observed halo-to-stellar mass relation,
thereby constituting another major problem of the particle dark matter
models.

Finally, SID is studied for systems falling in an external field,
assuming the field does not vary much across the system. This is the
case when a star cluster or a satellite dwarf galaxy orbits within a
much larger host field. The external field truncates the phantom dark
matter halo radius, and therewith reduces the mass of the phantom dark
matter halo. A $d(M_{b,\rm host})$ relation is considered for the
dwarf galaxies compiled by \citet{Miller+2013} assuming they are
embedded in an external field.  SID predicts that for these galaxies
there must be nearby gravitational sources like a companion galaxy or
a cluster of galaxies. For each deviating dwarf galaxy, the distance
to the external field source and the baryonic mass of the external
field source follows a simple relation (Eq. \ref{md}), shown in
Fig. \ref{host}. Also satellite galaxies near more massive hosts will
have larger radii on average as a result of the truncation of their
phantom dark matter halo masses due to the external field.  The
hitherto unexplained size difference between Andromeda and Milky Way
satellites may be thus perhaps resolved. This implies that the
gravitational forces acting between two galaxies vary with the
position and mass of a third galaxy. 

In summary, real galaxies follow SID/Milgromian dynamics rather than
Newtonian dynamics, and reality is thus properly described by
Milgromian dynamics \citep{Famaey_McGaugh2012,Kroupa2012}. Additional
tests of Milgromian dynamics on star-cluster scales and on
cosmological scales are required to further ascertain its range of
validity.

\section{Acknowledgments}
We thank Mordehai Milgrom for his helpful and important comments. 
We thank Stacy McGaugh for sharing his observational data of the
mass discrepancy and acceleration of disc galaxies. Part of this 
work is done in Argelander-Institut f\"{u}r Astronomie der 
Universit\"{a}t Bonn, and Xufen Wu gratefully acknowledges support 
through the Alexander von Humboldt Foundation.

\appendix

\section{Milgromian dynamics/MOND}
\label{MOND}

In this section a brief introduction to Milgromian dynamics (i.e. to
Modified Newtonian Dynamics, MOND), is provided. The reader is
referred to \cite{Famaey_McGaugh2012} for a deep and thorough review
of this topic. Essentially, MOND is the full classical description of
gravitational dynamics encompassing the low-acceleration pure-SID
(Sec.~\ref{STSI}) regime and the Newtonian regime.

In addition to the major problems found in $\Lambda$ cold dark matter
($\Lambda$CDM) cosmological simulations based on Newtonian/Einsteinian
gravity \citep[e.g. the cusp, the satellite over-prediction and the
disc-of satellites problems, see][]{Pawlowski+2014, Ibata+2014,
  Kroupa2014, Kroupa2012,Springel+2008,Klypin1999,Moore+1999}, it is
also difficult to explain some puzzling conspiracies of C/WDM and
baryons in galaxies by means of the current best $\Lambda$C/WDM models
\citep{Sanders2009,Kroupa+2010,Famaey_McGaugh2012}. For instance,
there is an empirical relation between observed galaxy baryonic mass,
i.e. the luminosity, and rotation speed. This is the baryonic
Tully-Fisher relation \citep[hereafter
BTFR,][]{Tully_Fisher1977,McGaugh+2000,McGaugh2012}, which cannot be
naturally obtained in $\Lambda$CDM models \citep{McGaugh2012}. Apart
from the BTFR, there are other newly observed coincidences which are
difficult to be reproduced in the simulated galaxies in the
dark-matter framework, such as the observationally-deduced apparent
dark matter content in the tidally formed dwarf galaxies
\citep{Gentile+2007b} and the discovery of a universal scale for the
surface density of both the baryons and dark matter halo at the core
radius of effective dark matter in a galaxy
\citep{Gentile+2009,Milgrom2009b}.  Indeed, the standard model of
cosmology is in poor agreement with data and the hypothesis that C/WDM
particles play a significant role in the universe has been seriously
challenged if not ruled out
\citep{Sanders2009,Kroupa+2010,Kroupa2012,Famaey_McGaugh2012,
  Kroupa2014}.

Milgromian dynamics was originally proposed
\citep{Milgrom1983a,Milgrom1983b,Milgrom1983c} to account for
gravitational dynamics in the classical regime without introducing
DM. With MOND, Milgrom extends our understanding of effective
gravitational dynamics beyond the gravitational-dynamical systems
known in~1916.  \citet{BM1984} demonstrated that MOND conserves
momentum, energy and angular momentum in a self-gravitating system. In
MOND the dynamical acceleration, $g=|{\bf g}| = \sqrt{g_{\rm N} a_0}$, takes
the place of Newtonian acceleration, $g_{\rm N}$, in the weak field regime
where $g\ll a_0$ and which is the regime of pure-SID
(Sec.~\ref{STSI}); while dynamical acceleration approaches Newtonian
acceleration in the strong field regime, i.e., $g = g_{\rm N}$ when $g\gg
a_0$. The constant acceleration, $a_0 \simeq 1.21 \times 10^{-10}
ms^{-2} \approx 3.8\,$pc/Myr$^2$, is the critical value of
acceleration which switches dynamics between Newtonian and Milgromian
\citep[e.g.,][]{Milgrom1983a,BM1984,Begeman+1991,Kent1987,Milgrom1988,
  McGaugh2011,
  McGaugh2012,Sanders_McGaugh2002,Bekenstein2006,Milgrom2008}.  It is
found to be in coincidence with various constants of cosmology, such
as $a_0\approx cH_0/2\pi$, $a_0\approx c(\Lambda/3)^{1/2}/2\pi$, where $c$
is the speed of light in vacuum, $H_0$ is the local Hubble constant
and $\Lambda$ is the cosmological constant
\citep{Milgrom1983a,Milgrom1989,Milgrom2009a,Milgrom2014a}.

MOND has until now passed all tests over a wide range of scales in
different types of galaxies and naturally accounts for the
aforementioned observations \citep{Milgrom1983a,BM1984,McGaugh2011,McGaugh2012, 
Milgrom_Sanders2003,Sanders_Noordermeer2007,Gentile+2007b,Gentile+2009,McGaugh2004,Milgrom2009b},
such as the BTFR and the apparent dark matter content in tidal dwarf
galaxies. Moreover, MOND is very successful in explaining the vertical
kinematics of stars in galactic discs in the absence of dark matter
\citep{Bienayme+2007} and naturally accounts for the faster rotational
speeds of polar rings \citep{Lughausen+2013}. Furthermore, there are
other new covariant theories equivalent to MOND at their
non-relativistic limit \citep{Bekenstein2004,Zlosnik+2007,Bruneton+2007a,Zhao2007,
Sanders2005,Skordis2008,Skordis_Zlosnik2012,Halle+2008,Milgrom2009c}.

\section{Virial masses calculalted based on Miller's empirical
  relation}\label{mess}

The virial masses of particle dark matter halo masses cannot be
measured directly. However, because the rotation curves of late-type
galaxies are about constant to large radii, the masses can be
estimated by measuring the rotation speed within the outer regions of
the luminous galaxy component, which corresponds to the inner halo
region (Sec.~\ref{CDMmasses}). Considering that the virial masses of
the haloes in \citet{Miller+2013} are converted from the inner
circular velocity, $V_{2.2}$, of the galaxies, we now apply the same
method to calculate the virial masses for the same galaxies in
MOND. The total stellar masses, $M_s$, and $V_{2.2}$ are known in
\citet{Miller+2013}, and they follow a simple fitting function
\citep{Miller+2011,Miller+2013} \footnote{The last term on the right
  hand side of Eq. 5 in \citet{Miller+2011} should not exist.},

\beq \label{msv22} 
\log_{10} \left(\frac{M_s}{\msun}\right)=[a+b\log_{10}
\left(\frac{V_{2.2}}{\kms}\right)]. 
\eeq 

\noindent
Here $a=0.57\pm 0.48$ in $M_s$ and $b=4.35\pm 0.62$. From
Eq. \ref{vc} we know that $M_{\rm b}=\frac{V_{200}^4}{Ga_0}$. Thus for
gas-poor galaxies, the relation is 

\beq \label{poor} 
\log_{10}\left(\frac{V_{200}^4}{Ga_0}\right)=a+b\log_{10} (V_{2.2}); 
\eeq 

\noindent while for gas-rich galaxies,
$M_s=M_{\rm b}/2=\frac{V_{200}^4}{2Ga_0}$, the relation is 

\beq \label{rich}
\log_{10} \left(\frac{V_{200}^4}{2Ga_0}\right)=a+b\log_{10} (V_{2.2}). 
\eeq

\noindent Combining with Eq. \ref{mv200}, the virial masses of the
galaxies in the sample of \cite{Miller+2013} are: 

\beq \label{mvirmond}
\log_{10} \left( \frac{\mvir}{h}\right)=
\frac{3}{4}[a+b\log_{10} (V_{2.2})+\log_{10} (a_0)],~~({\rm gas-poor}); \nonumber \\
\eeq \beq \log_{10} \left( \frac{\mvir}{h}\right)=\frac{3}{4}[a+b\log_{10}
(V_{2.2})+\log_{10} (2a_0)],~~({\rm gas-rich}).  
\eeq 

\noindent
However, the fitting parameters for the Tully-Fisher relation
introduced by \citet{Miller+2011,Miller+2013} are strongly
sample-dependent, and the errors for $V_{2.2}$ are large and up to
$60\%$ (see Table 1 in \citealt{Miller+2013}). Hence the virial masses
calculated with Eq. \ref{mvirmond} are unreliable.  Thus we do not
calculate the virial masses of the galaxies using Eq. \ref{mvir} and
we use Eq. \ref{mvirmond} instead.

\bibliographystyle{mn2e}
\bibliography{ref}
\end{document}